\documentclass[a4paper,dvipdfmx]{article}
\usepackage{graphics}
\usepackage{shonan}
\usepackage{color}
\usepackage{cite}
\usepackage{url}

\begin{document}

\SHONANno{222}
\SHONANtitle{The Future of Development Environments with AI Foundation Models}
\SHONANauthor{%
Xing Hu\\
Raula Gaikovina Kula\\
Christoph Treude}
\SHONANdate{October 6--9, 2025}
\SHONANmakecover

\title{The Future of Development Environments with AI Foundation Models}
\author{Organizers:\\
Xing Hu\\
Raula Gaikovina Kula\\
Christoph Treude}
\date{October 6--9, 2025}
\maketitle

%
%
%
%

\section*{Background and Introduction}
Recently, AI Foundation models (e.g., LLMs like GPT-4, ChatGPT\footnote{\url{https://chatgpt.com/}} Copilot \footnote{\url{https://copilot.microsoft.com/}}, and Code models like llama2\footnote{\url{https://www.llama.com/models/llama-3/}} and Deepseek\footnote{\url{https://www.deepseek.com/en}}) have attracted attention from both academia and industry. In Software Engineering, several studies showed that these models can achieve remarkable performance in various tasks, e.g., code generation, testing, code review, and program repair \cite{Deng2024,Xia2023,Yuan2024,gao2025current,pearce2025asleep}. For example, Hu et al. explored the performance of AI FMs on five software engineering tasks, and they found that AI FMs outperform other state-of-the-art approaches in code generation, unit test case generation, and automated program repair by large margins . Xia et al. found that directly applying AI FM can already substantially outperform all existing APR techniques . Yuan et al. proposed the first ChatGPT-based unit test generation approach which improves the SOTA substantially.

Traditionally, the IDE (Integrated Development Environment) has placed source code as the key artifact, and used version control to help manage the project. With the introduction of Foundation Models, the models, data and the natural language used to query the AI FMs will need to be managed. Considering that AI FM such as GPT-4 was developed by AI researchers rather than SE researchers, we are unsure how software engineers interact with Foundational Models. Hence we raise three questions:

\begin{enumerate}
    \item What should the future IDE look like? How can we improve the experience of human-machine interaction? Do we still need to write the code in the future?
    \item What are the challenges and opportunities that Foundational Models bring into light? What are the potential risks and drawbacks associated with using AI FMs in software engineering, and how can they be mitigated?
    \item How can we build various AI FM agents into this new IDE?
\end{enumerate}

With respect to designing the IDE, constructing an FM is costly and time-consuming and dependent on the dataset. The high cost of training of FMs requires us to have new engineering techniques to support the training and application of FMs, i.e., LLM engineering. Also, an FM can be trained with different hardware and software platforms, and there are many open-source FMs, e.g., StarCoder [5]. However, it is still unknown how to implement FMs across different hardware and software platforms. In summary, FM engineering has the following challenges:

\begin{enumerate}
    \item What types of data are most suitable for training an FM, and what strategies should be employed for effective AI dataset management?
    \item What methods can be utilized to identify and rectify bugs in an FM?
    \item What are the strategies for effectively deploying FMs across diverse hardware and software platforms?
    \item Given the dependency of prompt design and use on the underlying FM and the varying performance of prompts across different FMs, how should prompts be approached in the context of software engineering? For example, should we consider developing a prompt language that aligns closely with our programming language?
\end{enumerate}

To understand the impact of GenAI on the Integrated Development Environment (IDE), 33 experts from the three domains of Software Engineering (SE), Artificial Intelligence (AI), and Human-Computer Interaction (HCI) gathered to discuss the grand challenges and opportunities at Shonan Meeting 222 ``\textit{The Future of Development Environments with AI Foundation Models}''\footnote{\url{https://shonan.nii.ac.jp/seminars/222/}}.
To help organize the Shonan, we sent out a pre-event survey that highlighted five questions that researchers felt that were important when discussing the future of development environments:
\begin{itemize}
\item Which software development tasks should GenAI handle?
\item How should GenAI be integrated into IDE features?
\item What role remains for humans in software development?
\item Do we still need development environments?
\item Your boldest claim for software engineering in 2050?
\end{itemize}

\clearpage

\section*{Meeting Schedule}

\begin{itemize}
\item Check-in Day: October 5th 2025 (Sun)
\begin{itemize}
\item Welcome Reception
\end{itemize}
\end{itemize}

\begin{itemize}
\item Day 1 : October 6th 2025 (Mon)
\begin{itemize}
\item Shonan Introductions
\item Attendee Visions (1 - 33)
\item Open Discussions on Breakouts (open)
\item Roadmaps and Feedback - current IDEs, abilities of the agents, outcomes for participants (open).
\end{itemize}
\end{itemize}

\begin{itemize}
\item Day 2 : October 7th 2025 (Tue)
\begin{itemize}
\item Morning Breakout 1 - breakup into evolution, agents \& prompts, 2050, humans and process. 
\item Morning Breakout 2 - walk and talk sessions
\end{itemize}
\end{itemize}

\begin{itemize}
\item Day 3 : October 8th 2025 (Wed)
\begin{itemize}
\item Morning Breakouts 3 - deciding on evolution, agents \& prompts, 2050, humans and process subgroups.
\item Breakouts feedback (open).
\item excursion and banquet
\end{itemize}
\end{itemize}

\begin{itemize}
\item Day 4 : October 9th 2025 (Thur)
\begin{itemize}
\item Morning reflections and documentation (breakout)
\item after morning break closing and discussion of future collaborations (open).
\end{itemize}
\end{itemize}

\clearpage

\section*{List of Participants}
\begin{itemize}
  \item Dr.~Xing Hu – Zhejiang University, China
  \item Dr.~Raula Gaikovina Kula – The University of Osaka, Japan
  \item Dr.~Christoph Treude – Singapore Management University
  \item Dr.~Xin Xia – Zhejiang University, China
  \item Prof.~Sebastian Baltes – Heidelberg University, Germany
  \item Prof.~Denys Poshyvanyk – William \& Mary, USA
  \item Prof.~Shinpei Hayashi – Institute of Science Tokyo, Japan
  \item Prof.~Daniel German – University of Victoria, Canada
  \item Prof.~Zhongxin Liu – Zhejiang University, China
  \item Dr.~Jin Guo – McGill University, Canada
  \item Prof.~Marco Gerosa – Northern Arizona University
  \item Dr.~Igor Steinmacher – Northern Arizona University
  \item Dr.~Mairieli Wessel – Radboud University, The Netherlands
  \item Dr.~Markus Wagner – Monash University, Australia
  \item Prof.~Marc Cheong – University of Melbourne, Australia
  \item Prof.~Michele Lanza – Università della Svizzera italiana, Switzerland
  \item Prof.~Reid Holmes – University of British Columbia, Canada
  \item Prof.~Fabio Calefato – University of Bari, Italy
  \item Prof.~Nicole Novielli – University of Bari, Italy
  \item Prof.~Sonia Haiduc – Florida State University, USA
  \item Prof.~Takashi Kobayashi – Institute of Science Tokyo, Japan
  \item Ms.~Annie Vella – Westpac NZ, NewZealand
  \item Prof.~Junjie Chen – Tianjin University, China
  \item Prof.~Robert Hirschfeld – Hasso Plattner Institut, University of Potsdam
  \item Prof.~Earl Barr – University College London, UK
  \item Dr.~Shinobu Saito – NTT, Japan 
  \item Prof.~Kazumasa Shimari – NAIST, Japan
  \item Prof.~Olivier Nourry – The University of Osaka, Japan
  \item Prof.~Youmei Fan – NAIST, Japan
  \item Prof.~Kelly Blincoe – University of Auckland, New Zealand
  \item Prof.~Yintong Huo – Singapore Management University, Singapore
  \item Prof.~Laurie Williams – North Carolina State University, USA
  \item Prof.~Andrian Marcus – George Mason University, USA
\end{itemize}

\clearpage

\section*{Day 1 - Introductions and Attendee Visions}
The first day started with the introductions of the Shonan meeting and organizers. 
Then we moved to each of the 33 participants to introduce their research and answer the following questions that were formulated from a pre-survey from the participants themselves.
The questions were:

\begin{itemize}
\item Which software development tasks should AI handle?
\item How should AI be integrated into IDE features?
\item What role remains for humans in software development?
\item Do we still need development environments?
\item Your boldest claim for software engineering in 2050?
\end{itemize}

These questions were derived from the following themes of software tasks, IDE features, the human component of IDEs and more radical ideas of the future.

At the end of the day, participants had discussions of the how to proceed with these topics, as there was disagreement of whether to take stock of the current trends and existing GenAI tools and trends that were driven by the industry, the AI, SE and HCI communities and how the IDE falls between the different domains. 
Participants were asked to place themselves on an axis of social vs technical, short-term to long-term projects. 

\section*{Day 2 - Breakout groups formation}

The first session included an open discussion on whether the meeting should break up into smaller groups or try to work on a broader perspective. 
After much discussion, it was decided that there would be four breakout groups to focus on different perspectives on the future of IDEs.
At the same time, the group decided that there would still be an overlapping theme that all participants should still aim for.
The four groups were
\begin{itemize}
    \item Breakout Group 1 - A discussion on the existing tools such as LLM models, agents and other tooling that has already made its way into existing IDEs. The claim is that the IDE will undergo a revolution change to how we build software.
    \item Breakout Group 2 - A discussion on how the AI tools are just the next step in the evolution of Software development. The technology will empower how we build software. 
    \item Breakout Group 3 - A discussion on the human and process integration into the IDE. The group discussion takes into account different viewpoints and how the human process may change in this new IDE.
    \item Breakout Group 4 - A discussion on the radical ideas of the IDE. This discussion is about how we re-imagine the IDE based on the promises of GenAI. Almost all discussion is mainly science fiction and stretches the limits of the possibilities of the IDE.  
\end{itemize}

The rest of the afternoon was spent in deep discussions and then a walking session with team members. 

\section*{Day 2 to 4 - Breakout groups discussions}
\SHONANabstract{Breakout group: 1 }{%
\paragraph{Participants:} 
Daniel German,
Marco Gerosa,
Shinpei Hayashi,
Yintong Huo,
Takashi Kobayashi,
Michele Lanza
Andrian Marcus,
Olivier Nourry,
Denys Poshyvanyk,
Christoph Treude,
Annie Vella,
Xin Xia,
}

\paragraph{Context:}
Some researchers view AI-augmented IDEs as an evolutionary step rather than a radical overhaul of software engineering practice. The core tasks—design, coding, debugging, testing—remain unchanged; GenAI simply automates the repetitive or mechanical aspects, freeing developers to devote more time to high-level design and creative problem-solving. This automation is expected to streamline workflows, reduce boilerplate work, and accelerate delivery cycles, thereby enhancing overall productivity.

Despite these benefits, significant hurdles persist. Aligning training data across the software engineering and AI domains demands careful curation, while divergent terminologies—such as symbolic versus neural representations—require interdisciplinary dialogue to create shared ontologies. Moreover, GenAI is increasingly taking care of low-level technicalities and idiosyncrasies, pushing abstraction layers higher within IDEs. This shift promises greater developer autonomy but also risks obscuring underlying implementation details if not managed thoughtfully.

Ultimately, programming must remain an immersive, collaborative human activity. As GenAI handles more routine tasks, developers will be left to intervene only during rare yet critical moments—debugging complex issues or architecting scalable systems, as recent research on developer-AI collaboration shows how interaction patterns are shifting. Ensuring that automation enhances rather than diminishes creativity requires IDE designs that encourage intentional interaction and maintain the developer's central role in decision-making.

\paragraph{Claims:} 
\textbf{The key claim is that although the technologies are changing, the basic components of software development and its key guiding principles remain the same. 
We are simply moving to a higher level of abstraction, and all tedious work will be handled by the computing power, leaving tasks of critical thinking. }

\SHONANabstract{Breakout group: 2}{%
\paragraph{Participants: } Reid Holmes, Kelly Blincoe, Laurie Williams, Sebastian Baltes, Earl Barr, Kazumasa Shimari, Zhongxin Liu, Junjie Chen, Shinobu Saito}

\paragraph{Context:}

Although GenAI represents an exciting avenue for broad improvement in how we build software, past technological innovations have also greatly impacted (and improved) how we build systems today.
For example, the emergence of high-level languages and compilers obviated the need for engineers to work directly with low-level source code (and even to examine it at all!). 
Rather than this significant advancement, which greatly reduced the mechanical effort required to create and evolve software, reducing the size of software engineering teams, it instead greatly increased their capability to tackle large problems leading to increasing software demand.
Similar observations can be made for Integrated Development Environments, which expanded access to tooling and reliable debugging, open-source software and reusable frameworks which made composing systems more practical, and automated debugging and continuous integration approaches which made the evaluation and validation of systems both more practical and repeatable.
Rather than decimating the field of Software Engineering, we believe GenAI will continue this trend of increasing the overall capacity of software development teams, enabling them to tackle more impactful projects than ever before.

\paragraph{Summary of Discussion:} 
Our initial discussions revealed consistent confusion about the differentiation between evolution and revolution in this space.
By carefully examining available evidence and past technological advancements that have pushed the field of software engineering forward, we reached consensus that evolution was more appropriate than revolution, even if the breadth of the tasks GenAI impacts is larger than many past advances.

To sharpen our discussion and focus on concrete ways that GenAI pushes software engineering forward without ending the profession as we know it, we focused on five commonly-held misconceptions both from popular media and academic literature and discussed their practical impact on the field. These five themes included:

\begin{itemize}
    \item \textbf{GenAI will replace all Software Engineers.} This prediction has been made for many past advances in Software Engineering, such as compilers, open source libraries, and IDEs. In each instance, these advances simply made it more efficient for engineers to build larger systems than ever before. While GenAI may be more expansive than any one of these improvements, we believe that the tools will ultimately improve the capabilities of human engineers, rather than wholesale replacing them.
    
    \item \textbf{Prompts are the new source code.} Currently, source code is the only representation of a system that can deterministically create the same program every time. Until this is resolved, or GenAI tools can prove that isomorphic programs are equivalent, source code will remain the dominant underlying representation of software systems.
    
    \item \textbf{Vibe-coding is the only process.} Vibe-coding approaches have shown great promise for prototypes and proof-of-concept of work, but it remains to be seen whether  these approaches will translate to systems with strict performance, privacy, and security constraints.
   
    \item \textbf{Greenfield generalizes to brownfield.} While GenAI-based approaches have shown great strengths automating tasks within new systems, the incredible size and complexity of established systems will continue to strain these new approaches in different ways not experienced during Greenfield development.
  
    \item \textbf{All you need is another agent.} While agents will continue to automate specific tasks, fundamental uncertainty during software development will necessitate human involvement coordinating many of these activities.
\end{itemize}

\paragraph{Key Claim:} 
\textbf{The primary claim of our three-day discussion is that GenAI represents a continued evolution of the tools, techniques, and practices Software Engineers use to create the modern systems we rely upon every day.}

\SHONANabstract{Breakout group: 3}{%
\paragraph{Participants:} Youmei Fan,
Sonia Haiduc,
Marc Cheong,
Nicole Novielli,
Fabio Calefato,
Daniel German,
Igor Steinmacher,
Robert Hirschfeld, and
Andrian (Andi) Marcus} 

\paragraph{Context:}
Our vision for the future of human and process integration in IDEs emphasizes context-aware collaboration between developers and AI agents that dynamically adapt to individual personas, project types, and emotional states. 

Future IDEs will move beyond treating all users as generic programmers -- in a `one-size-fits-all' approach -- and instead serve as personalized cognitive companions. For example, \textit{domain experts} (such as scientists, teachers, or physicians) or any \textit{layperson} who needs to develop software to solve personal problems/tasks (e.g., analyzing a dataset) or for their personal use will interact with IDEs that translate intent into finished software (or intermediate code?), offering contextual explanations rather than low-level syntax guidance. \textit{Professional developers} will utilize environments that prioritize workflow efficiency, automated documentation, and process transparency to minimize disruptions within teams. Professionals using IDEs as part of their job in \textit{mission-critical organizations building software for clients} will rely on environments/IDEs that integrate process governance, knowledge management, and privacy-preserving collaboration across distributed contributors. In all cases, the IDE becomes adaptive, learning from user preferences, workflow/usage history, criticality of the project, and documentation artifacts to automate repetitive processes. This will shape the IDE with different plugins, agents, feedback, and workflows around each persona’s context and emotional state. 

Furthermore, documentation in future IDEs will move beyond static project instructions or API references to prompt-recording and intent-recording. As developers interact with IDEs through natural language, the IDE will capture and organize these prompts, responses, and contextual decisions as part of the project’s documentation. This will allow both humans and machines to trace not only the resulting code but also the reasoning, alternatives, and discussions that shaped it. Such ``prompt-as-documentation'' will make development more transparent, auditable, and learnable, transforming documentation from artifact into a component of process memory and collaboration.

\paragraph{Challenges}

However, realizing this vision poses challenges. First, we foresee a need to balance automation and human autonomy, ensuring that AI tools empower rather than replace human developer agency and responsibilities.  Building transparency, trust, and security into the collaboration is key to avoiding over-reliance on non-transparent recommendations, keeping accountability, and decision-making process in a reliable level. It may also be challenging to create an environment in which it is possible to keep the productivity and quality without disrupting team coordination or introducing cognitive overload by having excessive automation. 

More broadly, the emergence of these adaptive/persona-driven environments has the potential to introduce concerns related to over-personalization, privacy, and bias. Collecting behavioral and affective data to tailor experiences risks negatively impacting user consent and autonomy, and may inadvertently privilege certain persona profiles over others. Developers may fear that continuous sensing and modeling will become a form of \textit{surveillance capitalism}, discouraging creativity or self-expression. From this same perspective, as AI systems increasingly mediate decision-making, there is a risk that critical reasoning, craftsmanship, and accountability will diminish in favor of algorithmic efficiency.

Future research should focus on developing ways to represent persona-driven interactions between humans and AI, defining ethical boundaries for adaptive process automation, and empirically assessing the impact on productivity, satisfaction, and collaboration. Ultimately, we envision IDEs evolving into socio-technical ecosystems that empower human capabilities that define ``modern'' software engineering, fostering a sustainable human-AI relationship.

\paragraph{Key Claim:} 
\textbf{Future IDEs should evolve into adaptive, context-aware cognitive companions that integrate human and process aspects by dynamically adapting to users’ personas, project contexts, and emotional states—thus transforming software development into a personalized, human-AI collaborative ecosystem.}

\SHONANabstract{Breakout group: 4}{%
\paragraph{Participants:} Markus Wagner, Jin Guo, Mairieli Wessel, Xing Hu, Raula Kula, 
Robert Hirschfeld, Marc Cheong, and Michele Lanza}

\paragraph{Context:}

This `futuring' exercise imagines the context of programming in the future. 
The point of this exercise is to report about what we can hypothetically `see' in 2050 -- from the perspective of the further future, in the spirit of \textit{futurecasting} -- without being too concerned on `how to get there'.
We pretend we are in the year 3050, uncovering an ancient software/system development environment from 2050. We dig up parts and try to understand what we see. 

\paragraph{Immersion, Time Travel, and Translation}

By walking in the door, the ancient system turns on filling in the room. However, it seems to clearly distinguish itself from is surroundings, yet be immersive enough for the user/creator to be familiar with the functions. Looks like the developers might have reached a consensus on how invasive the system is to the user/creator needs. 

\paragraph{Artifacts}
The artifacts in 2050 are equipped with AIs and multi-modals. Among these AIs, there is a super AI that serves as the role of CPU.
The super AI controls the running of the artifact.
By integrating the AIs, the artifacts can evolve and fix bugs themselves without lines of code from human.
Considering the ``compiler'', it could detect ambiguities in natural languages.
In addition, the compiler is designed for AIs to make them run without syntax errors.

\paragraph{Software development \textit{versus} usage}
The distinction between creating and using software no longer exists. People generate and modify software components directly through natural language, gestures, or biological and contextual signals. Every interaction with a system can alter its behaviour, so users effectively become developers. The development environment interprets intent across multiple forms of input, allowing software to adapt instantly to new needs. Programming is not a separate activity but a shared process between human and system, unfolding continuously through everyday use rather than within discrete development phases.

\paragraph{Simulator}
Taking a leaf from the sci-fi franchises of Star Trek and Warhammer 40k, the holodeck metaphor applies to the IDE of the future, enabling “wireless communion with the machine spirit”.
What-if scenarios such as “What if Google Maps routes everybody down the same road?” can be answered using full-scape, full-sized digital twins. 
In a boon for agility and rapid prototyping, dials for “fidelity” and “scope” allow rapid re-evaluation of the current scenario.
Edge cases and exceptions, as well as unintended changes over time, are best represented as a “focus on `clusters of unique outcomes'” as what is “unintended” will be revealed upfront if the system actually records everything. Motivated by shifts/advances in quantum computing, many states can be explored in parallel, in a world where ‘compute power’ is not a problem, and abstractions and jumps are possible when needed. 
Multiple users (or developers) can engage with the system together, with tools in place to enable one to go into the system and to make changes: this blurs the boundaries between “creator vs user”.

\paragraph{Feedback Loop}
From observing the ``unintended outcome'' of the software changes in the simulator to having a clear understanding of how to steer the program toward more intended ones, an effective feedback loop is essential. As the system becomes complex, this feedback loop increasingly becomes diagnostic to help developers reason the immediate next step. Moreover, as the boundaries of using and programming software blur, the level of expertise in developing software diversifies more extensively. Therefore, the granularity and focus of the feedback are automatically adjusted to be appropriate for the developers' technical background. It is important to mention that the effect observed from the impact simulator is both social and technical; each of them is situated on different dimensions that are interconnected. To enable the calibration, the feedback loop is multidimensional that encourages debate and reflection on complex social and value-laden issues. The link between social and technical dimensions is explicit to facilitate translating them into concrete guidelines for technical implementation. Finally, the feedback loop provides mechanisms for coordination and establishing norms to resolve conflicts and debates raised from different parties involved during the technology advancement.

\paragraph{Key Claim:} 
\textbf{The key lesson of the exercise and claim is that it might be time to let go of conventional notions of files and folder organization, version control, and source code as the primary artifact. A future beyond the keyboard as the main mode of input is possible, while collaboration between humans remains foundational to IDE. }

\section*{Collaborations and future directions} 

By the end of the workshop, some of the breakout groups decided to submit separate articles on the their respective breakouts. 
For example, group 3 plan to finalize an initial draft of an article that takes the position that GenAI represents a broad evolutionary step in how we build software systems. We plan to submit this article to a venue like Communications of the ACM, and also hope that another breakout session will take the counterpoint position that GenAI is fundamentally reshaping the field. The goal of these pieces are to provide insight into persistent confusion in the field about GenAI and its role in Software Engineering.
As such, details of the discussions were saved for the upcoming publications.

The report is a broad summary of the discussions, with more publications to come. 
The next step as mentioned above is to further express the four different perspectives of the workshop. 
We plan to publish in both shortform as short abstracts and those that would like to also publish in longform such as a book. We are planning to submit a proposal to Shonan to publish a book on this topic.
As researchers, our role is to provide awareness and more pushback on the trends happening in Software Engineering. 

We hope these articles will spur the kinds of important conversations we currently need in the community as we work to build common ground for our discussions that balance GenAI hype with the industrial reality of large-scale software systems.
\clearpage

\bibliographystyle{unsrt}
\bibliography{references}

\end{document}